\documentclass[%
 reprint,
 twocolumn,
 superscriptaddress,
 showpacs,
 preprintnumbers,
 nofootinbib,
 amsmath,
 amssymb,
 aps,
 prd,
 floatfix,
]{revtex4-1}
\usepackage[american]{babel}
\usepackage{epsfig}
\usepackage{amsfonts,amsthm,graphicx,psfrag}
\usepackage{hyperref}

\newcommand{\beq}{\begin{eqnarray}}
\newcommand{\eeq}{\end{eqnarray}}

\newcommand{\tr}{\operatorname{Tr}}

\newcommand{\roundcustom}{\operatorname{round}}

\begin{document}
\preprint{}

\title{Lattice study of thermodynamic properties of dense QC$_2$D}

\author{N.~Astrakhantsev}
\email[]{nikita.astrakhantsev@physik.uzh.ch}
\affiliation{ Physik-Institut, Universit{\" a}t Z{\" u}rich, Winterthurerstrasse 190, CH-8057 Z{\" u}rich, Switzerland }
\affiliation{ Institute for Theoretical and Experimental Physics NRC ``Kurchatov Institute'', Moscow, 117218 Russia }

\author{V.\,V.~Braguta}
\email[]{braguta@itep.ru}
\affiliation{ Institute for Theoretical and Experimental Physics NRC ``Kurchatov Institute'', Moscow, 117218 Russia }
\affiliation{ Moscow Institute of Physics and Technology, Dolgoprudny, 141700 Russia }
\affiliation{ Bogoliubov Laboratory of Theoretical Physics, Joint Institute for Nuclear Research, Dubna, 141980 Russia } 
\affiliation{ National University of Science and Technology MISIS, Leninsky Prospect 4, Moscow, 119049 Russia }

\author{E.-M.~Ilgenfritz}
\email[]{ilgenfri@theor.jinr.ru}
\affiliation{ Bogoliubov Laboratory of Theoretical Physics, Joint Institute for Nuclear Research, Dubna, 141980 Russia }

\author{A.\,Yu.~Kotov}
\email[]{kotov@itep.ru}
\affiliation{ Institute for Theoretical and Experimental Physics NRC ``Kurchatov Institute'', Moscow, 117218 Russia }
\affiliation{ Bogoliubov Laboratory of Theoretical Physics, Joint Institute for Nuclear Research, Dubna, 141980 Russia }
\affiliation{ National University of Science and Technology MISIS, Leninsky Prospect 4, Moscow, 119049 Russia }

\author{A.\,A.~Nikolaev}
\email[]{aleksandr.nikolaev@swansea.ac.uk}
\affiliation{ Department of Physics, College of Science, Swansea University, Swansea SA2 8PP, United Kingdom }

\begin{abstract}
In this paper we study thermodynamic properties of dense cold $SU(2)$ QCD within lattice simulation with dynamical rooted staggered quarks which in the continuum limit correspond to $N_f=2$ quark flavours. We calculate baryon density, renormalized chiral and diquark condensates for various baryon chemical potentials in the region $\mu \in (0,\,2000)$~MeV. It is found, that in the region $\mu \in (0,\,540)$~MeV the system is well described by the ChPT predictions. In the region $\mu > 540$~MeV the system becomes sufficiently dense and ChPT is no longer applicable to describe lattice data. For chemical potentials $\mu > 900$~MeV we observe formation of the Fermi sphere, and the system is similar to the one described by the Bardeen-Cooper-Schrieffer theory where the the diquarks play a role of Cooper pairs. In order to study how nonzero baryon density influences the gluon background we calculate chromoelectric and chromomagnetic fields, as well as the topological susceptibility. We find that the chromoelectric field and the topological susceptibility decrease, whereas the chromomagnetic field increases with rising of baryon chemical potential. Finally we study the equation of state of dense two-color quark matter.
\end{abstract}


\maketitle

\section{Introduction}

Study of Quantum Chromodynamics at finite baryon density is an important research topic of modern physics which is closely connected to various problems in astrophysics and cosmology. Experimental studies of QCD at finite baryon density can be carried out in heavy ion collision experiments. In particular, the region of the phase diagram with high temperature and small baryon density is well explored at the Large Hadron Collider (LHC) and Relativistic Heavy Ion Collider (RHIC), while the physical programs of future Facility for Antiproton and Ion Research (FAIR) and Nuclotron-based Ion Collider Facility (NICA) are focused on large baryon density and small temperature. 

At the moment, the theoretical understanding of the QCD phase diagram in $(\mu, T)$ plane is rather schematic, since the most powerful approach, lattice simulation of QCD, cannot be directly applied in the region of finite density due to the sign problem~\cite{Muroya:2003qs}. Numerous lattice attempts to overcome the sign problem provide reliable information only in the region of small baryon density~\cite{Ratti:2019tvj}. In the absence of straightforward results from lattice simulation of QCD, one applies different analytical approaches to study the 
$(\mu, T)$ phase diagram: mean field approaches~\cite{Alford:1998mk, Alford:1999pa, Buballa:2003qv, Khunjua:2017mkc, Khunjua:2018sro, Khunjua:2020vrp, Khaidukov:2019icg}, the method of Dyson-Schwinger equations and the renormalization group~\cite{Fischer:2018sdj,Fu:2019hdw, Gao:2020qsj, Son:1998uk}, the large--$N_c$ approach~\cite{McLerran:2007qj}, perturbative QCD~\cite{Gorda:2014vga, Gorda:2018gpy} and others.
Although the results obtained theoretically are important, it is rather difficult to estimate the reliability of these predictions. 

One of the possible ways to shed light on the properties of dense media is
to apply lattice simulation to theories which are similar to QCD but are not plagued by the sign problem. Although such theories differ from real QCD in some aspects, it is believed that these QCD-like theories can provide important information common for dense media in general. The most popular choices are the QCD at finite isospin density~\cite{Son:2000xc, Kogut:2002zg, Kogut:2002tm, Brandt:2017oyy, Brandt:2018wkp} and the two-color QCD at finite baryon density~\cite{Kogut:1999iv,Kogut:2000ek}. This paper is devoted 
to lattice study of the dense two-color QCD. 

The two-color QCD at finite chemical potential has been studied with lattice simulations quite intensively, see, {\it e.g.}~\cite{Hands:1999md, Kogut:2001if, Kogut:2002cm, Muroya:2002ry, Cotter:2012mb, Braguta:2016cpw, Holicki:2017psk, Boz:2018crd, Wilhelm:2019fvp, Iida:2019rah, Boz:2019enj, Buividovich:2020dks} and references therein. Mostly, these papers are aiming at the study of the phase diagram of two-color QCD in the region of small and moderate baryon densities. 

The phase structure of dense two-color QCD and its properties were studied in our previous papers~\cite{Bornyakov:2017txe, Astrakhantsev:2018uzd, Bornyakov:2019jfz,  Bornyakov:2020kyz}, where lattice simulations were carried out at a relatively small lattice spacing $a=0.044$\,fm. In this paper we also employ this spacing. Compared to previous studies at larger lattice spacings, this allows us to extend the range of accessible values of the baryon density, up to quark chemical potential $\mu>2000$\,MeV, while avoiding strong lattice artifacts.

In this paper we are going to continue these studies. In particular, we shall calculate chiral condensate, diquark condensate and quark number density for various values of chemical potential under investigation. In order to study how nonzero baryon density influences the properties of the gluon background we calculate chromoelectric, chromomagnetic fields and the topological susceptibility. 
In addition, we shall study the equation of state of dense two-color QCD. 

The manuscript is organized as follows. In the next section we describe our lattice set-up and details of the calculation of the observables under consideration. In section~\ref{sec:fermion} we present our results on fermionic observables. In section~\ref{sec:gluon} we study how nonzero baryon density modifies the properties of the gluon background. Section~\ref{sec:gluon} is devoted to our study of the equation of state of dense two-color QCD. Finally, in the last section we discuss our results and draw the conclusion. 

\section{Details of the calculation}
\label{sec:simdet}

In this section we briefly describe our lattice setup. More details can be found in papers~\cite{Braguta:2016cpw, Astrakhantsev:2018uzd}. In our lattice study we used the tree level improved Symanzik gauge action~\cite{Weisz:1982zw, Curci:1983an}.
For the fermionic degrees of freedom we used staggered fermions with an action of the form
\begin{equation}
\label{eq:S_F}
S_F = \sum_{x, y} \bar \psi_x M(\mu, m)_{x, y} \psi_y + \frac{\lambda}{2} \sum_{x} \left( \psi_x^T \tau_2 \psi_x + \bar \psi_x \tau_2 \bar \psi_x^T \right),
\end{equation}
where $\bar \psi$, $\psi$ are staggered fermion fields, $M(\mu, m)_{x, y}$ is the standard Dirac operator for  staggered fermions, $m$ is the bare quark mass. The chemical potential $\mu$ is introduced into the Dirac operator through the multiplication of the links along
and opposite to the temporal direction by factors $e^{\pm \mu a}$, respectively. 

In addition to the standard staggered fermion action we add a diquark source term~\cite{Hands:1999md}
to equation (\ref{eq:S_F}). The diquark source term explicitly violates $U_V(1)$ and allows to observe diquark 
condensation even on finite lattices, because  this term effectively chooses one vacuum from the family of $U_V(1)$-symmetric vacua. 

In the present study we are going to investigate a theory with the partition function
\beq
Z=\int DU e^{-S_G} \cdot {\bigl ( \det (M^\dagger M + \lambda^2) \bigr )}^{\frac 1 4},
\label{z2}
\eeq
where $S_G$ is the tree level improved Symanzik gauge action. In the continuum, the action (\ref{z2}) corresponds to $N_f=2$ dynamical fermions.

The results presented in this paper have been obtained in lattice simulations performed on a $32^4$ lattice for the set of chemical potential values in the region $a\mu \in (0, 0.5)$. In out simulation we use the
quark mass $am = 0.0075$ corresponding to the pion mass $m_{\pi}=741(15) \mathrm{\,MeV}$ ($a m_\pi = 0.165 \pm 0.003$). At zero density we performed the scale setting using the QCD Sommer scale $r_0=0.468(4) \mathrm{\,fm}$\cite{Bazavov:2011nk}. In this case the string tension associated to $\mu=0$ amounts to $\sqrt{\sigma_0}=476(5) \mathrm{\,MeV}$ at $a = 0.044 \mathrm{\,fm}$. 

Typically, one carries out numerical simulations at a few nonzero values of $\lambda$ and then extrapolates to zero $\lambda$. 
Notice, however, that numerical simulations in the region of large baryon density are numerically very expensive. For this reason, in this paper we have chosen a different strategy. Most of our lattice simulations are conducted at a single fixed value $\lambda=0.00075$ or $\lambda/m\simeq 0.1$. However, in order to check the $\lambda$-dependence of our results for chemical potentials $a\mu=0.0, 0.1, 0.2, 0.3, 0.4$ we carry out additional lattice simulations at $\lambda=0.0005, 0.001$. We found that the calculations of most observables at $\lambda/m\simeq 0.1$ give results close to $\lambda \to 0$ extrapolation and our conclusions are not effected by finite value of the $\lambda$ parameter (see discussion below).

\begin{figure}[b]
    \centering
    \includegraphics[scale=0.55]{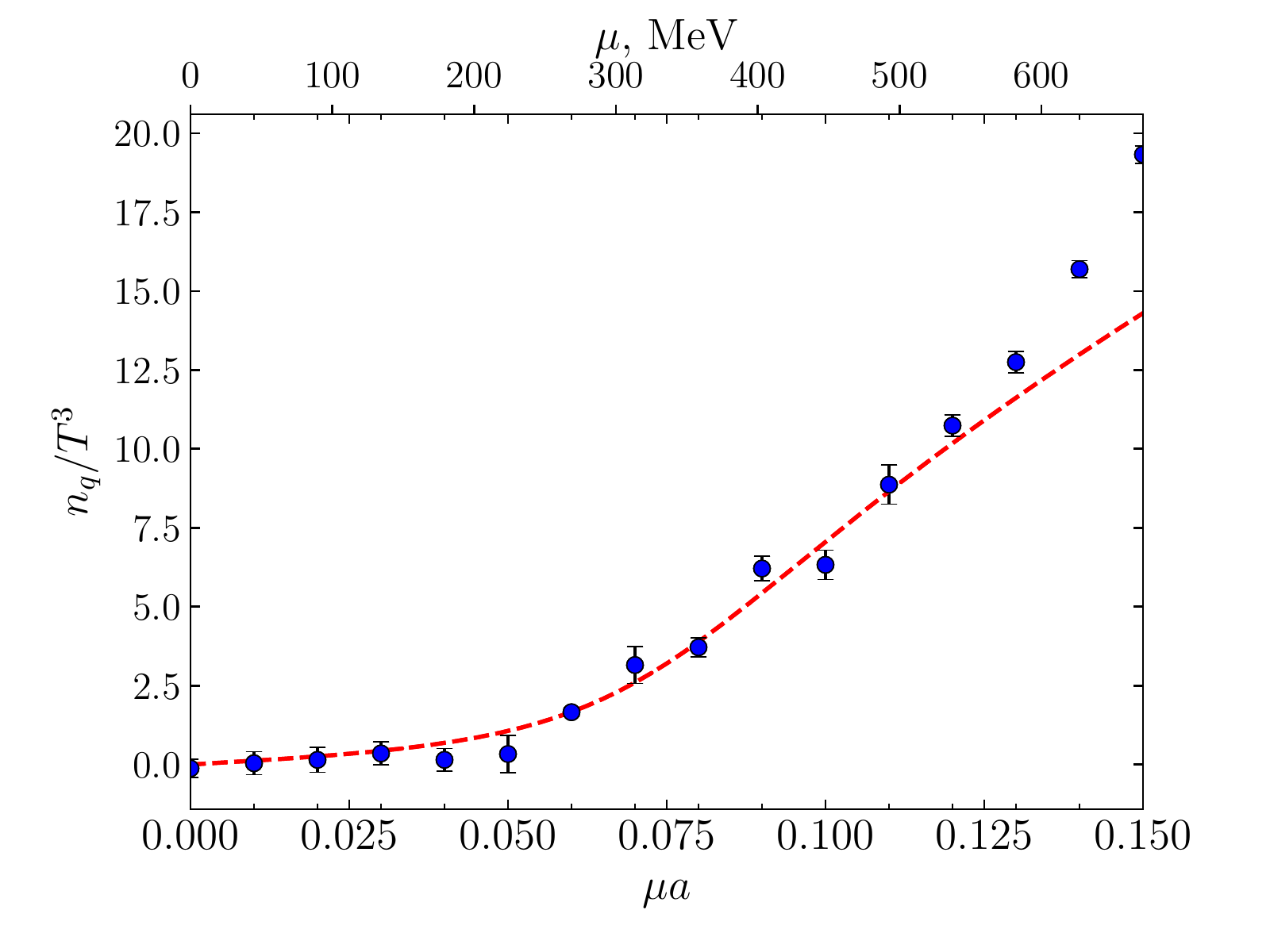}
    \caption{Quark number density as a function of chemical potential. Dashed red line represents the fit by (\ref{eq:quark_density}), for detailed discussion see the paragraph after eq. (\ref{eq:G_constant}).}
    \label{fig:quark_dens}
\end{figure}

In our paper we are going to calculate the following fermionic observables
\begin{itemize}
\item The diquark condensate:
\begin{equation}
\label{eq:diquark_condensate1}
a^3 \langle qq\rangle=  \frac{1}{N_s^3 N_t} \frac{\partial (\log\,Z)}{\partial \lambda}\,,
\end{equation}
\item The chiral condensate:
\begin{equation}
\label{eq:chiral_condensate}
a^3\langle \bar q  q \rangle = a^3\langle\bar q_{i \alpha}  q_{i \alpha}\rangle =  \frac{1}{N_s^3 N_t}\frac{\partial (\log\,Z)}{\partial (ma)}\,;
\end{equation}
\item The quark number density:
\begin{equation}
\label{eq:bar_number}
a^3 n_q =  \frac{1}{N^3_s N_t} \frac{\partial (\log\,Z)}{\partial (\mu a)}\,;
\end{equation}
\end{itemize}
The baryon density is a conserved quantity and it does not require renormalization. The chiral and diquark condensates require renormalization. To this end, we are going to follow the renormalization procedure analagous to the paper~\cite{Brandt:2017oyy} where the authors studied QCD at nonzero isospin density\footnote{If instead of the pion condensate and the pionic source term in QCD at finite isospin density~\cite{Brandt:2017oyy} one considers the diquark condensate and the diquark source term in dense two-color QCD, both theories look similar in their properties.}:
\beq
\label{chiral_cond}
\Sigma_{\bar q q }= \frac {m} {4m_{\pi}^2 F^2 } \bigl [ 
\langle \bar q q \rangle_{\mu} - \langle \bar q q \rangle_{0}
\bigr ] + 1\\
\label{diquark_cond}
\Sigma_{q q}= \frac {m} {4 m_{\pi}^2 F^2 } \bigl [ 
\langle  q q \rangle_{\mu} - \langle q q \rangle_{0}
\bigr ]
\eeq

\begin{figure}[t]
    \centering
    \includegraphics[scale=0.55]{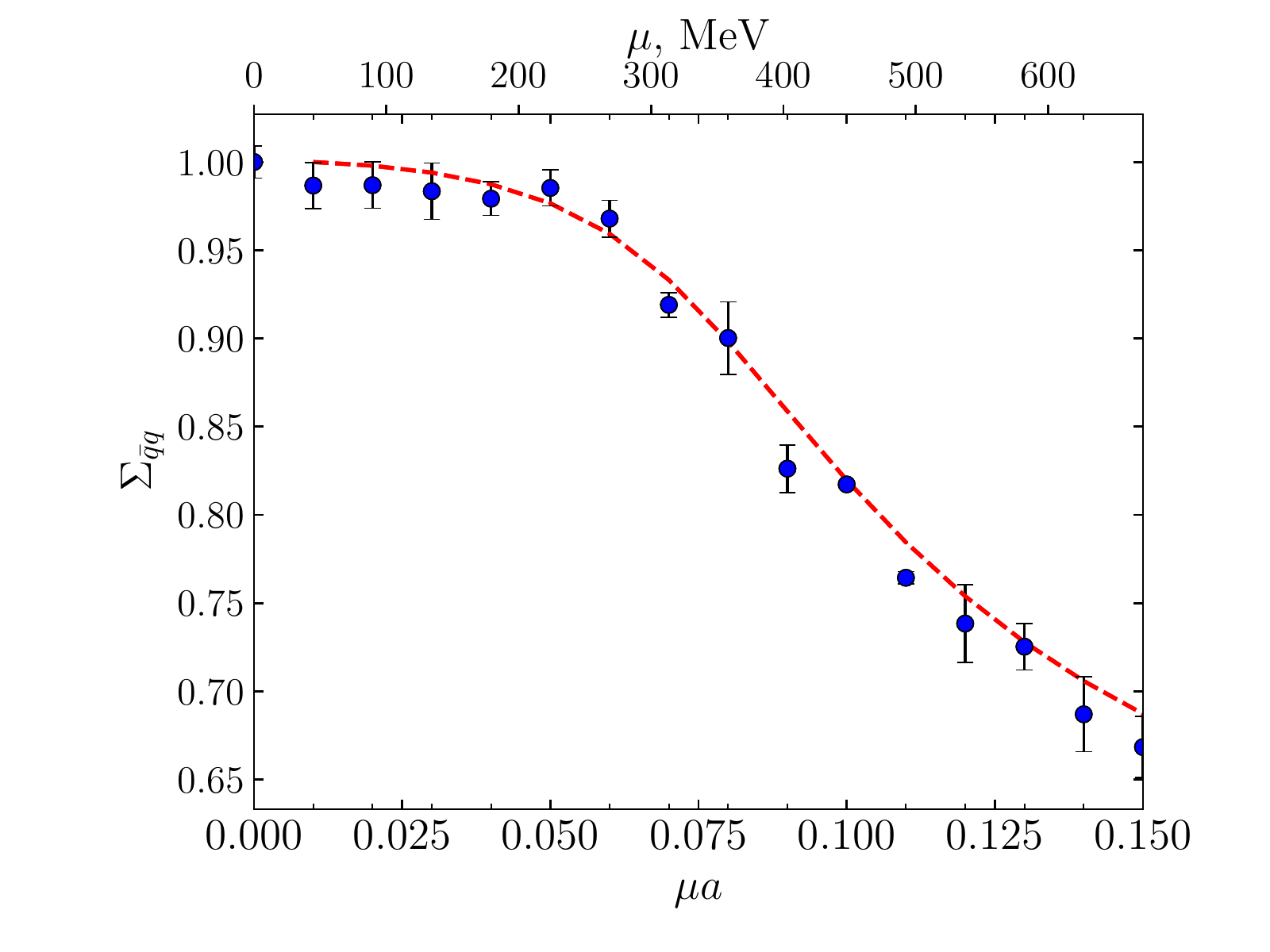}
    \caption{Renormalized chiral condensate (\ref{chiral_cond}) as a function of chemical potential. Dashed red line represents the fit by (\ref{eq:chiral_condensate_chpt}), for detailed discussion see the paragraph after eq. (\ref{eq:G_constant}).}
    \label{fig:ren_chiral_cond}
\end{figure}

To get rid of the additive divergences in the chiral and diquark condensates we subtract the corresponding observable $\langle \bar q q \rangle_{0}$ and $\langle q q \rangle_{0}$ at zero $\mu=0$. Since the quark mass is renormalized multiplicatively, $m_r=Zm$, multiplicative divergence falls out in the combination $m_r\partial/\partial m_r=m\partial/\partial m$ and the quantity $m(\langle \bar q q \rangle-\langle \bar q q \rangle_{0})$ has a well-defined continuum limit. Other factors in the formula (\ref{chiral_cond}) are introduced in such way, that the chiral condensate $\Sigma_{\bar q q }$ is $1$ at $\mu=0$ and $0$ if the chiral symmetry is fully restored. To do it  we also use the Gell-Mann--Oaks--Renner relation: $m^2_{\pi}=m \langle \bar q q \rangle / 4F^2$~\cite{Kogut:2000ek}, where $F$ is the constant in front of the kinetic term of the Chiral Perturbation Theory, 
in the leading order the pion decay constant $f_{\pi}=F/2$. 

One can easily check that at zero density $\mu=0$ the staggered Dirac operator $M(\mu=0,m)$ has the following form $M(\mu=0,m)_{x,y}=Q_{x,y}+ma\delta_{x,y}$, where the operator $Q$ satisfies $Q^{\dag}=-Q$. The partition function ($\ref{z2}$) at zero $\mu=0$ takes the following form: $Z=\int DU e^{-S_G} \cdot {\bigl ( \det (Q^{\dag}Q + (ma)^2+ \lambda^2) \bigr )}^{\frac 1 4}$ and depends on the quark mass and $\lambda$ only via the factor $m^2a^2+\lambda^2$. Thus, the diquark source parameter $\lambda$ and the quark mass $ma$ are completely equivalent at zero baryon density and their multiplicative renormalization factors coincide. As the consequence, the multiplicative renormalization of the diquark condensate (\ref{diquark_cond}) can be taken to be equal to the multiplicative renormalization factor of the chiral condensate (\ref{chiral_cond}).

\begin{figure}[t!]
    \centering
    \includegraphics[scale=0.55]{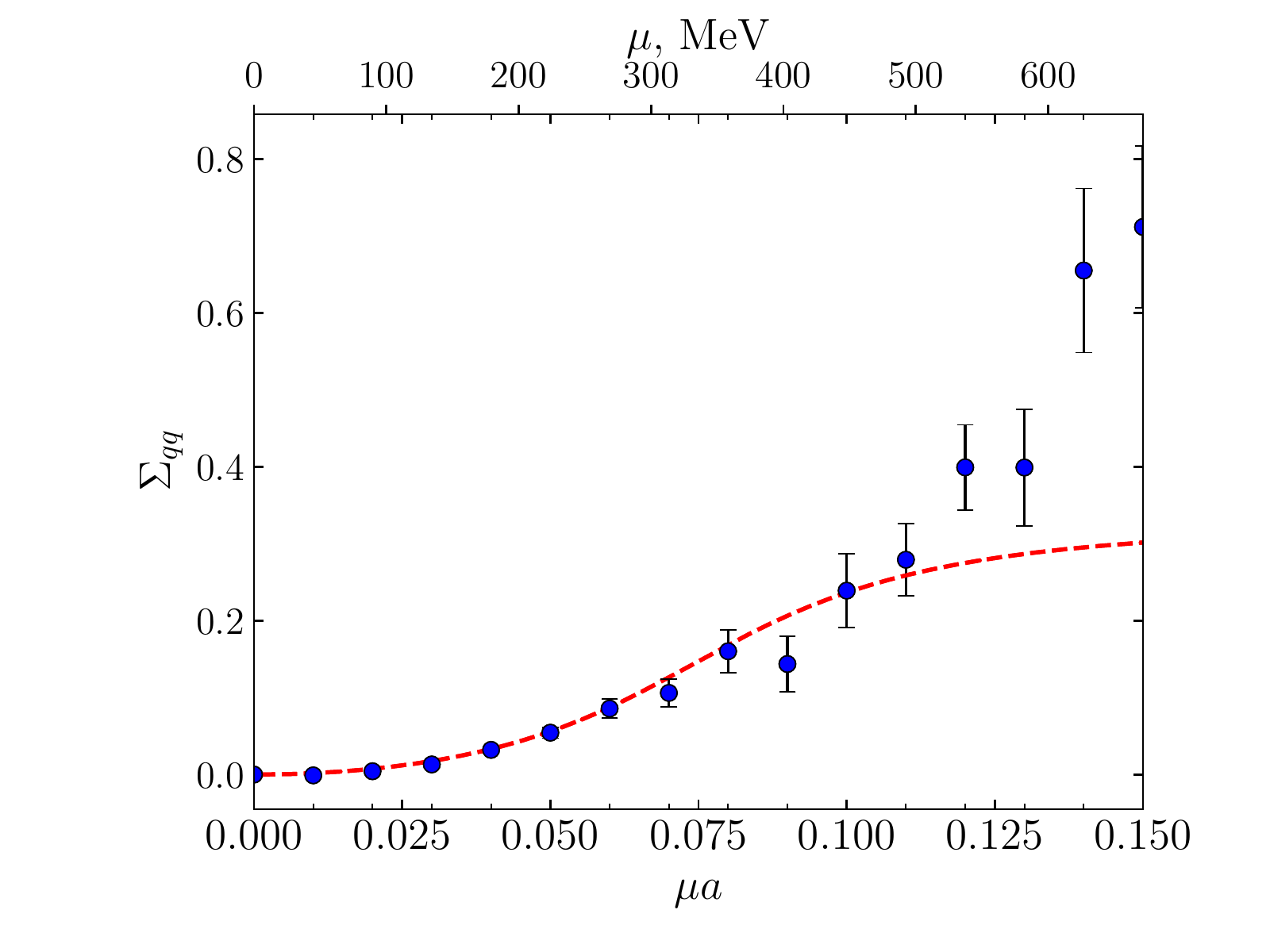}
    \caption{Renormalized diquark condensate (\ref{diquark_cond}) as a function of chemical potential. Dashed red line represents the fit by (\ref{eq:diquark_condensate}), for detailed discussion see the paragraph after eq. (\ref{eq:G_constant}).}
    \label{fig:ren_diquak_cond}
\end{figure}

To calculate the renormalized chiral and 
diquark condensates one needs to know the value of the constant $F$. To find this constant we fit our lattice results for the quark number density by the ChPT formula~(\ref{eq:quark_density}) in the region $a\mu \in (0,0.12)$ with the pion mass $m_{\pi} = 738 \pm 13$\,MeV ($a m_\pi = 0.165 \pm 0.003$). The fit quality is good $\chi^2/\mbox{dof} \sim 1$ and the fitting parameters are $F=60.8\pm1.6$\,MeV ($aF=0.01359 \pm 0.00036$).

In addition to the fermionic observables (\ref{eq:diquark_condensate1}), (\ref{eq:bar_number}), we study how nonzero density modifies the gluonic background. In particular, the following observables are calculated
\beq
\label{chromoelectric}
\frac {\langle \delta (\vec E^a)^2 \rangle} {T^4} = \frac {12 N_t^4} {g^2} \biggl ( 
\langle U_P^t \rangle_{\mu=0} - \langle U_P^t \rangle_{\mu}
\biggr ), \\
\label{chromomagnetic}
\frac {\langle \delta (\vec H^a)^2 \rangle} {T^4} = \frac {12 N_t^4} {g^2} \biggl ( 
\langle U_P^s \rangle_{\mu=0} - \langle U_P^s \rangle_{\mu}
\biggr ),
\eeq
where $U_P^s, U_P^t$ are the spatial and temporal plaquetes. It is clear that these observables show how the chromoelectric and chromomagnetic fields are affected by the baryon density.   

To study the topological properties of the dense two-color QCD we are going to calculate the topological susceptibility. 
The details for these measurements mainly follow~\cite{Bonati:2014tqa}. To get the topological charge on each configuration we use the Gradient Flow technique~\cite{Luscher:2009eq,Luscher:2010iy}. On the smoothened configurations we measure:
\begin{equation}
    Q_L=-\frac{1}{512\pi^2}\sum_x\sum_{\mu\nu\rho\sigma=\pm1}^{\pm4}\tilde{\epsilon}_{\mu\nu\rho\sigma}\tr U_{\mu\nu}(x)U_{\rho\sigma}(x)\,,
    \end{equation}
where $U_{\mu\nu}(x)$ is the plaquette at the point $x$ in directions $\mu$ and $\nu$. The final estimator for the topological charge $Q$ is given by:
\begin{equation}
    Q=\roundcustom\left(\alpha Q_L\right),
\end{equation}
where $\roundcustom$ gives the closest integer to its argument and the factor $\alpha$ is chosen in such a way that it minimizes 
\begin{equation}
    \langle \left(\alpha Q_L-\roundcustom\left(\alpha Q_L\right)\right)^2 \rangle\,.
\end{equation}

\begin{figure}[b]
    \centering
    \includegraphics[scale=0.55]{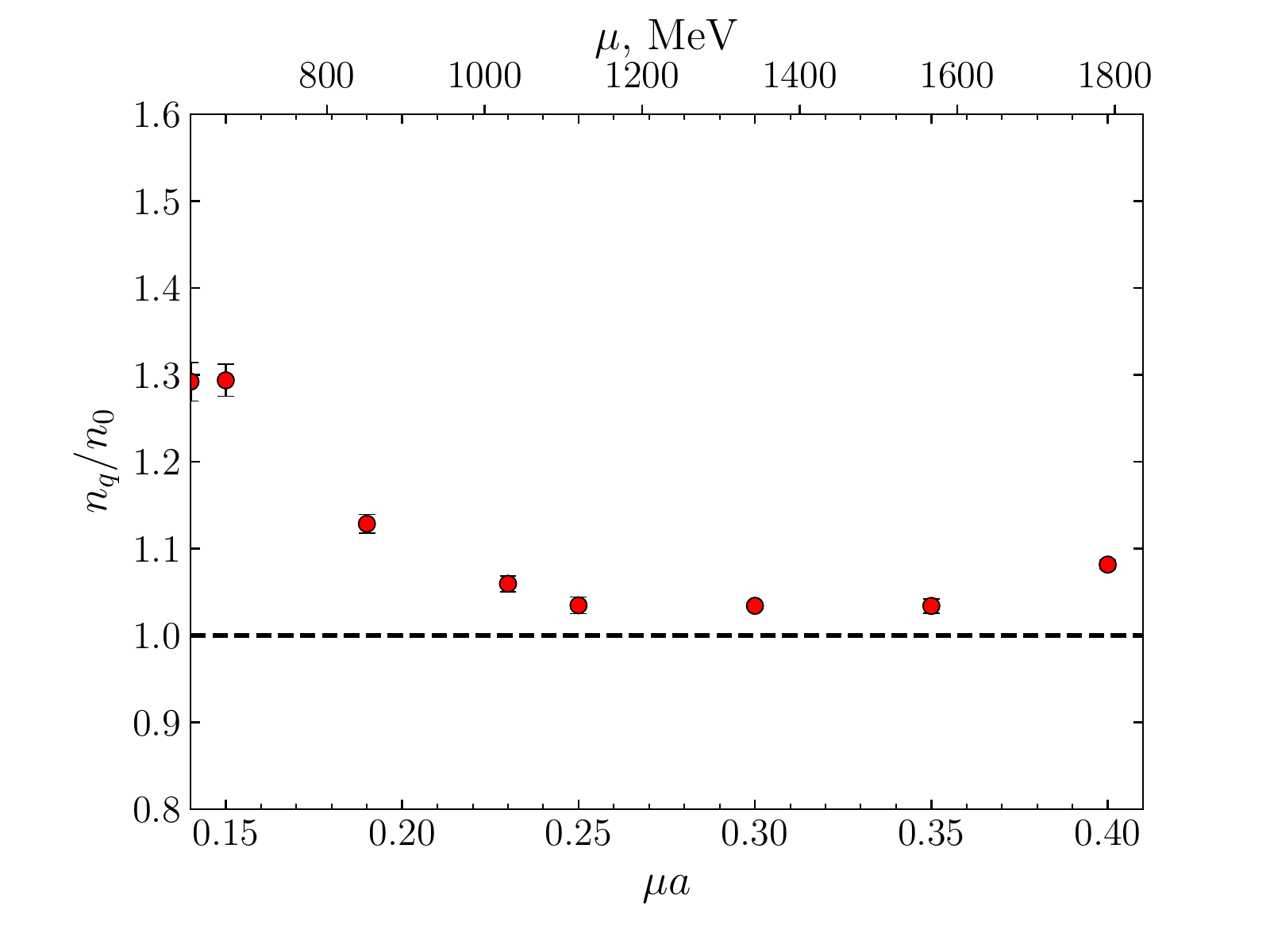}
    \caption{The ratio $n_q/n_0$ as a function of chemical potential, where $n_0=4\mu^3/3\pi^2$ is the quark number density for free relativistic quarks.}
    \label{fig:ratio_qnd}
\end{figure}

\begin{figure}[b]
    \centering
    \includegraphics[scale=0.6]{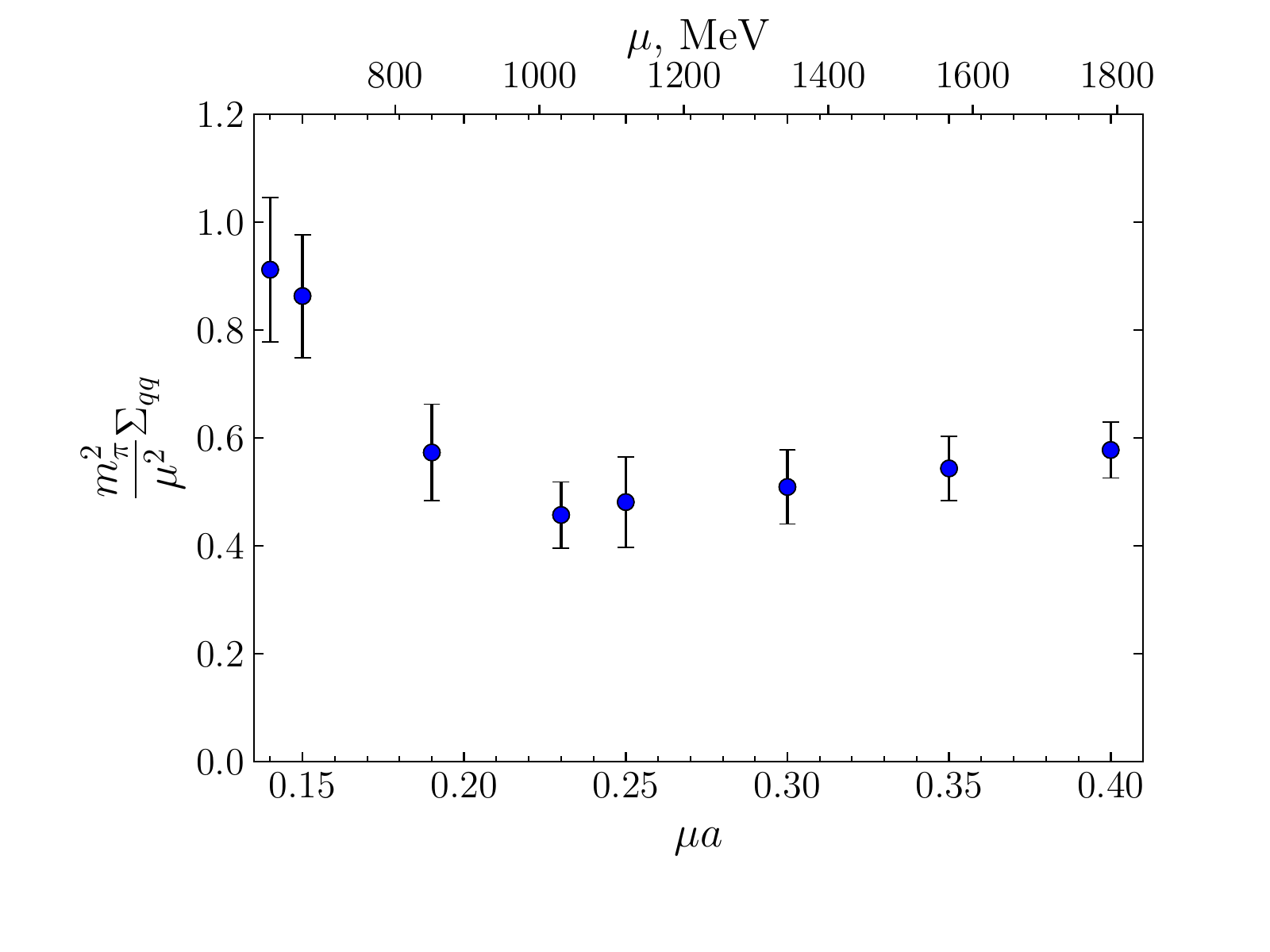}
    \caption{The ratio $m_{\pi}^2 \Sigma_{qq}/\mu^2$ as a function of chemical potential, where $\Sigma_{qq}$ is defined in (\ref{diquark_cond}).}
    \label{fig:ratio_dqc}
\end{figure}

By doing this we rescale the topological charge $Q_L$ so that its peaks become closer to integer values and then round the estimation to this integer, thus reducing lattice artifacts. The topological susceptibility is then given by
\begin{equation}
    a^4\chi_{\mbox{\footnotesize top}}=\frac{\langle Q^2 \rangle}{N_s^3N_{\tau}}\,.
\end{equation}

Finally we are going to study the equation of state (EoS) for the dense two-color QCD. The pressure $p$ can be calculated if one takes the integral of the quark number density
\beq
p(\mu)=\int_0^\mu d\xi n_q(\xi)+p(0).
\label{pressure}
\eeq
In this paper we work at zero temperature, where it is reasonable to take $p(0)=0$. In the calculation, the baryon density was interpolated by cubic splines. The pressure was then obtained with numerical integration of the interpolated baryon density.

Next let us consider the trace anomaly $I(\mu)= \langle T^{\mu}_{\mu} \rangle =\epsilon(\mu) - 3 p(\mu)$ which can be written as~\cite{Borsanyi:2010cj, Borsanyi:2012cr}
\beq
\label{anomaly}
&&\frac {I(\mu)} {T^4}= \frac {I_G(\mu)} {T^4} + \frac {I_F(\mu)} {T^4} \\
\label{eq:I_G}
&&\frac {I_G(\mu)} {T^4} = N_t^4
 \beta(g) \bigl [\langle S_G \rangle_{\mu} - \langle S_G \rangle_{0} \bigr ]\,, \\
\label{eq:I_F}
&&\frac {I_F(\mu)} {T^4} =-  N_t^4
\gamma(g)  ma \bigl [ \langle \bar q q  \rangle_{\mu}-\langle \bar q q  \rangle_{0}\bigr ]\,,
\label{eq:IF}
\eeq
where $I_G(\mu)$, $I_F(\mu)$ denote the gluon and fermion contributions to the anomaly,
$\langle S_G \rangle$ is the averaged value of the tree level improved Symanzik gauge action, and $\beta(g)$ and
$\gamma(g)$ are
\beq
\beta(g)=4\frac {d g^{-2}} {d \log a}, \\
\frac {d \log (ma)} { d \log a}= \gamma(g). 
\eeq
Here two comments are in order. 

First, in Eq.~(\ref{anomaly}) we subtracted the trace anomaly at zero temperature and density in order to get rid of the additive divergence.
Second, all simulations are carried out on the lattice $32^4$. Although one cannot exclude finite temperature or finite $N_t$ effects, the temperature in our simulations is close to zero. When we divide some observable by $T$ in the corresponding power (similar to formulae (\ref{chromoelectric}--\ref{chromomagnetic}) or (\ref{anomaly}--\ref{eq:IF}) ), instead of $T$ we imply inverse temporal size of our lattice $1/aN_t \simeq 140$\,MeV.

Having calculated the pressure and the trace anomaly we can calculate the energy density $\epsilon$ and the entropy density $s$
\beq
\label{eq:energy_density}
\epsilon &=& I + 3 p \\
\label{eq:entropy}
s &=& \frac {\epsilon + p - \mu n_q} T
\eeq

\section{Fermionic observables}
\label{sec:fermion}

The phase diagram of dense two-color QCD was studied previously within the ChPT in papers~\cite{Kogut:1999iv, Kogut:2000ek}. This phase diagram at zero $\lambda$ can be described as follows: for sufficiently small chemical potential the chiral symmetry is broken and the chiral condensate takes nonzero values, while the diquark condensate and baryon density are zero; at $\mu_c=m_{\pi}/2$ the system undergoes a second order phase transition where the diquark condensate plays a role of the order parameter; in the region $\mu > \mu_c$ the chiral condensate and the baryon density become nonzero. In the lattice formulation due to the finite pion mass the chiral condensate is nonzero after the transition, but it decreases with chemical potential. Moreover, nonzero values of the $\lambda$ parameter change the second order phase transition to a crossover. 
At the leading order approximation of the ChPT, the dependence of the diquark condensate, chiral condensate and quark number density on the chemical potential can be described by the following formulae~\cite{Kogut:2000ek}
\beq
\label{eq:diquark_condensate}
\langle q q \rangle  = 2 N_f G \sin \alpha\,,  \\ 
\label{eq:chiral_condensate_chpt}
\langle \bar q q \rangle  = 2 N_f G \cos \alpha\,,   \\ 
\label{eq:quark_density}
n = 8 N_f F^2 \mu \sin^2 \alpha\,,
\eeq
where the $\alpha$ angle can be extracted from the equation
\beq
\mu^2 \sin \alpha \cos \alpha = \mu_c^2 \biggl ( \sin \alpha - \frac {\lambda} {m} \cos \alpha \biggr )\,,
\eeq
and the constant
\beq \label{eq:G_constant}
G=\frac 1 {2 N_f} \sqrt { \langle \bar q q \rangle^2 + \langle q q \rangle^2 }\,.
\eeq
Our lattice results for the renormalized diquark condensate $\Sigma_{q q }$, the chiral condensate $\Sigma_{\bar q q }$ and the quark number density are presented in Fig.\,\ref{fig:ren_diquak_cond}, \ref{fig:ren_chiral_cond}, \ref{fig:quark_dens} correspondingly.

To proceed we fit simultaneously our lattice data for the diquark condensate, the chiral condensate and the quark number density in the region $a\mu \in (0,\,0.12)$ by modified formulae (\ref{eq:diquark_condensate})-(\ref{eq:quark_density}). The modification consists in addition of one constant $c_1$ to the quark condensate and the constant $c_2$ to the chiral condensate, these constants are aimed at account of the additive divergences which are contained in the lattice results. Thus in the fitting procedure we have three parameters: $F, \mu_c, c_1, c_2$. In Figs.\,\ref{fig:ren_diquak_cond}, \ref{fig:ren_chiral_cond}, \ref{fig:quark_dens} we present the results of this fit. From these figures it may be seen, that the fit quality is good ($\chi^2/\mbox{dof} \sim 1$) and $F = 63 \pm 3$\,MeV, $\mu_c = 403 \pm 30$\,MeV ($aF=0.0141 \pm 0.0006$), $a\mu_c=0.090 \pm 0.006$. The critical chemical potential $\mu_c$ obtained in the fitting procedure within the uncertainty agrees with that calculated from the pion mass: $\mu_c= m_{\pi}/2= 371 \pm 9$\,MeV. Since formulae above describe our lattice data quite well, we conclude that at low density ($\mu < 540$\,MeV) the system under study is well described by ChPT. A similar conclusion was also drawn in papers~\cite{Braguta:2016cpw, Iida:2019rah, Wilhelm:2019fvp}.

From Fig.~\ref{fig:ren_diquak_cond} and Fig.~\ref{fig:quark_dens} it may be observed that in the region $\mu > 540$\,MeV ($a\mu > 0.12$) the lattice data for the diquark condensate and quark number density start to deviate from the leading order ChPT predictions. To understand the origin of this deviation we remind, that at the leading order approximation of the chiral perturbation theory one may ignore the interactions between hadrons, what can be done when baryons form a dilute gas, where the interactions are not important. It is clear that the larger the baryon density the more important the interactions between baryons are, the larger the deviation from the leading order ChPT. Thus the deviation of lattice data from ChPT predictions can be considered as the transition of the system from a dilute baryon gas to dense matter phase. 

For sufficiently large density the wave functions of different baryons overlap. If the density is increased further, an individual quark no longer belongs to a particular baryon. One can expect that in this region of the chemical 
potentials, the system is similar in some properties to the Bardeen-Cooper-Schrieffer theory. Following~\cite{Cotter:2012mb} below we refer this region to as the BCS phase. In this phase the relevant degrees of freedom are quarks forming a Fermi sphere, and the baryon density is given by the one of non-interacting quarks $n_0 = 4 \mu^3 /3 \pi^2$. In other words, in the two-color QCD the diquarks are Cooper pairs in the BCS theory.

Note that the notion ``BCS phase'' is not fully appropriate to describe the system under study in the region of sufficiently large baryon densities. This is because the results of the BCS theory are applicable in the weak coupling regime, which might take place at ultrahigh densities only. In particular, one of the predictions of the BCS theory is $\Sigma_{qq} \sim \Delta(\mu) \mu^2$~\cite{Kanazawa:2009ks}, where the $\Delta(\mu)$ is the mass gap in the fermionic spectrum. In the weak coupling regime $\Delta(\mu) \sim \mu g^{-5} \exp {( - 3 \pi^2/\sqrt 2 g)}$~\cite{Son:1998uk}. But the baryon densities reached in our studies are moderate and the system in this region is still strongly coupled~\cite{Astrakhantsev:2018uzd}, thus the weak coupling formula for the mass gap is not applicable, while one might expect that the relation $\Sigma_{qq} \sim \Delta(\mu) \mu^2$ survives. The factor $\sim \mu^2$ in the last formula results from the fermionic density of states on the Fermi surface, and the factor $\Delta(\mu)$ determines the strength the $U_V(1)$ symmetry breaking in the system.

In order to find the value of the chemical potential where the BCS phase is formed, in Figs.~\ref{fig:ratio_qnd},~\ref{fig:ratio_dqc} we plot the ratios $n_q / n_0$ and $m_{\pi}^2 \Sigma_{qq}/\mu^2$, respectively. From Fig.~\ref{fig:ratio_qnd} it may be observed, that in the region $a\mu \in (0.2,\,0.4)$ the ratio $n / n_0$ deviates from the unity by not more than 10\,\%. What concerns the ratio $m_{\pi}^2 \Sigma_{qq}/\mu^2$, it goes to a plateau in the same region, {\it i.e.} the condensation of diquarks takes place on the surface of the Fermi sphere and the $\Delta(\mu)$ weakly depends on $\mu$. From these observations we can conclude, that in the region $a\mu > 0.2$ the system under study is in the BCS phase. The BCS phase in two-color QCD was observed previously in the papers~\cite{Cotter:2012mb,Braguta:2016cpw,Iida:2019rah,Boz:2019enj}.

\section{Gluonic observables}
\label{sec:gluon}

\begin{figure}[t]
    \centering
    \includegraphics[scale=0.525]{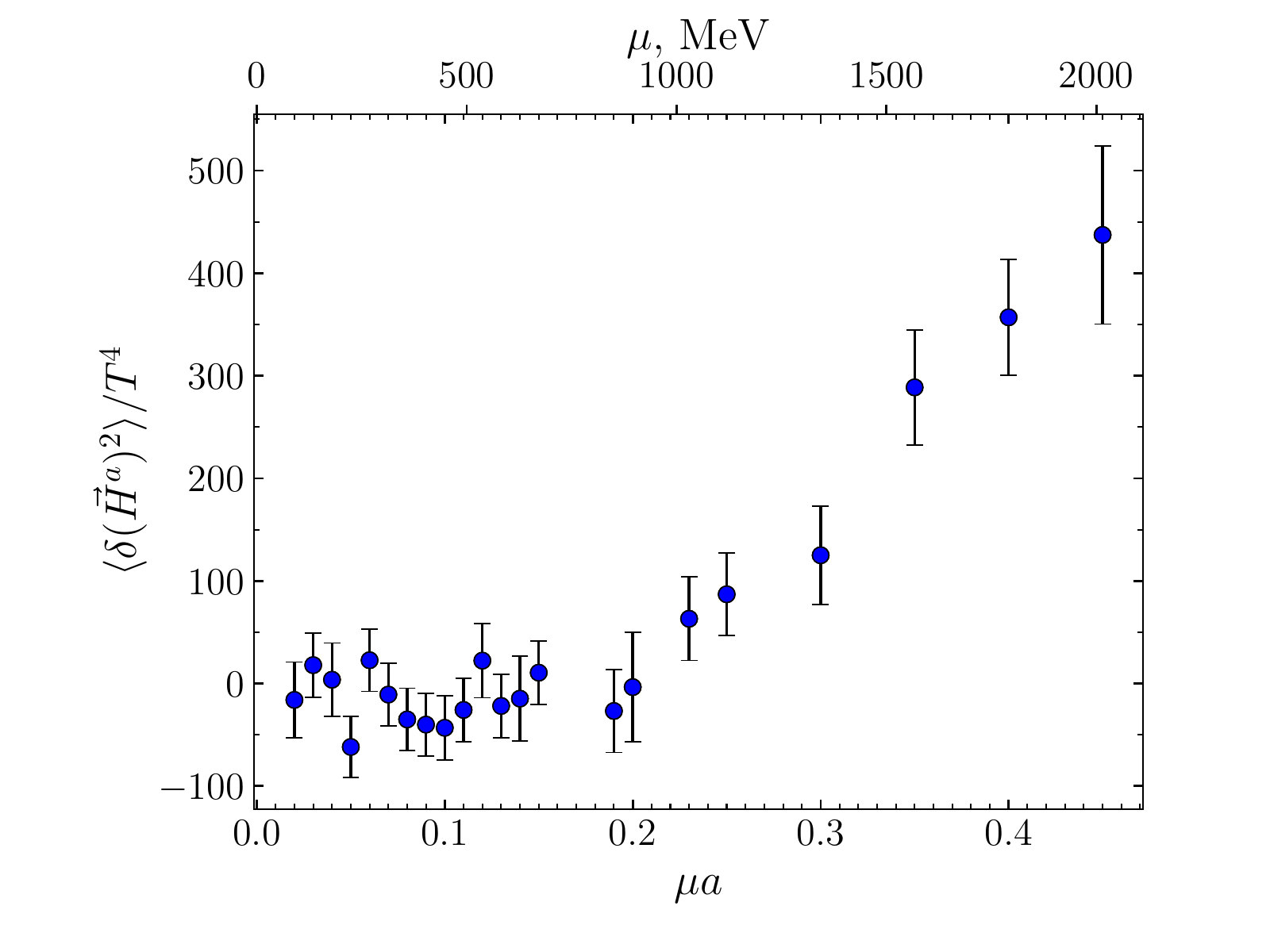}
    \caption{The ratio $\langle \delta (\vec H^a)^2 \rangle / T^4$, defined in (\ref{chromomagnetic}), as a function of chemical potential.}
    \label{fig:chromomagnetic}
\end{figure}

\begin{figure}[t]
    \centering
    \includegraphics[scale=0.525]{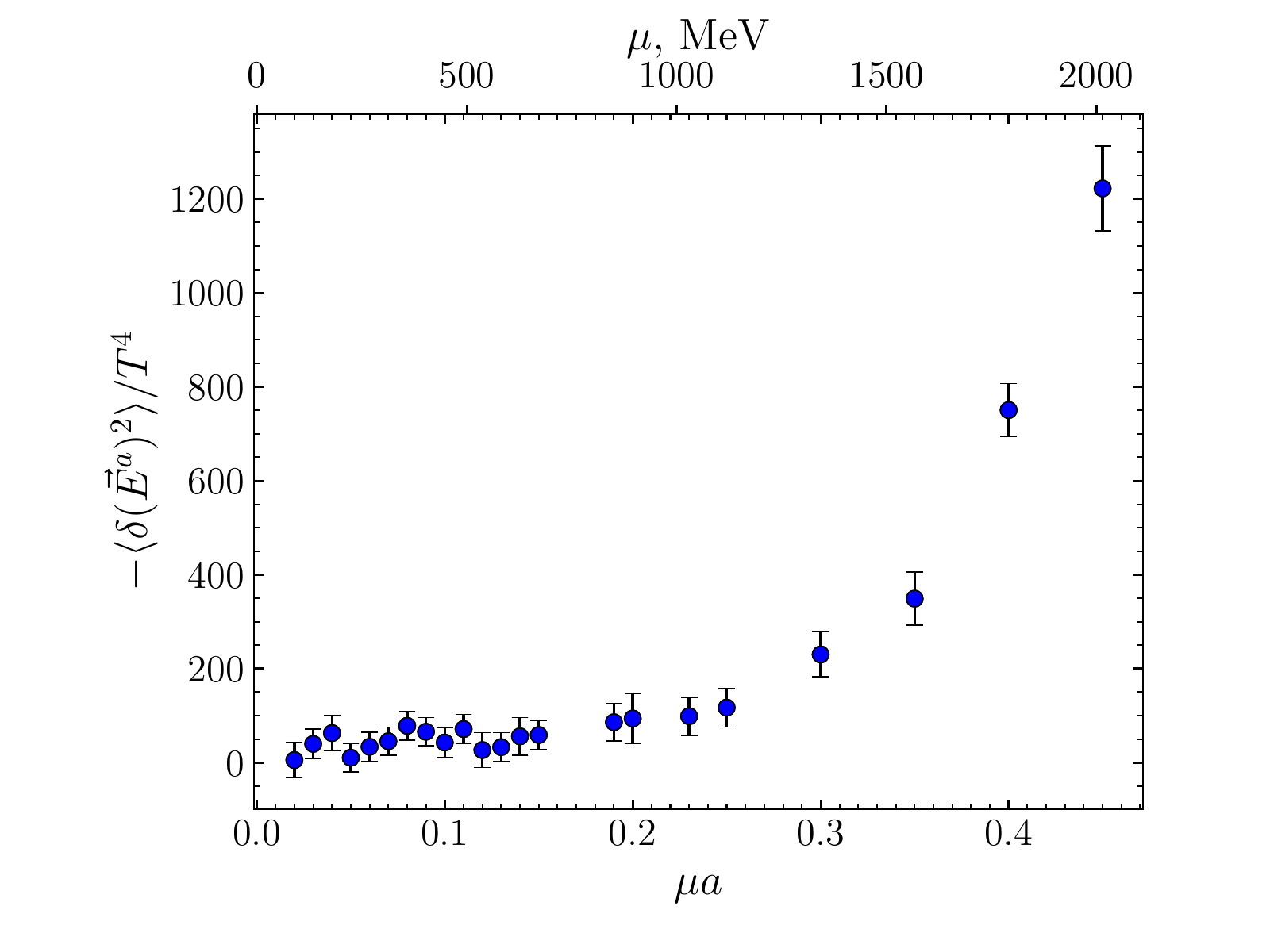}
    \caption{The ratio $-\langle \delta (\vec E^a)^2 \rangle/ T^4$, defined in (\ref{chromoelectric}), as a function of chemical potential. The minus sign is taken since the chromoelectric field decreases in dense matter as compared to the vacuum value.}
    \label{fig:chromoelectric}
\end{figure}

This section is devoted to the study of the gluon background at nonzero baryon density. To this end, in Fig.~\ref{fig:chromoelectric}, ~\ref{fig:chromomagnetic} we plot ratios (\ref{chromoelectric}), (\ref{chromomagnetic}) as functions of baryon chemical potential.

From Fig.~\ref{fig:chromoelectric} it may be observed, that the chromoelectric field decreases with increasing baryon density. We believe that this behaviour can be attributed to well known Debye screening of chromoelectric field in dense matter. This phenomenon was also observed in the study of Polyakov loop correlators~\cite{Astrakhantsev:2018uzd} and gluon propagators~\cite{Bornyakov:2019jfz, Bornyakov:2020kyz} in dense matter. It is interesting to note that in the BCS phase, chromoelectric field scales as $-\langle \delta (\vec E^a)^2 \rangle \sim \mu^4$.

Next let us consider the chromomagnetic field shown in Fig.~\ref{fig:chromomagnetic}. From this plot it is seen that within the uncertainty chromomagnetic field does not change as compared to its vacuum value up to $a\mu \sim 0.2$. In the region $a\mu > 0.2$ the magnetic field increases with rising of the baryon density. This behaviour can be explained if we recall that magnetic screening in QCD matter is related to nonperturbative spatial confinement. In the paper~\cite{Bornyakov:2017txe} it was shown, that in the region $a\mu > 0.2$ the spatial string tension decreases, {\it i.e.} spatial confinement plays a less important role, thus chromomagnetic field is less screened. Similar results were obtained in papers~\cite{Bornyakov:2019jfz, Bornyakov:2020kyz}.

\begin{figure}[b]
    \centering
    \includegraphics[scale=0.6]{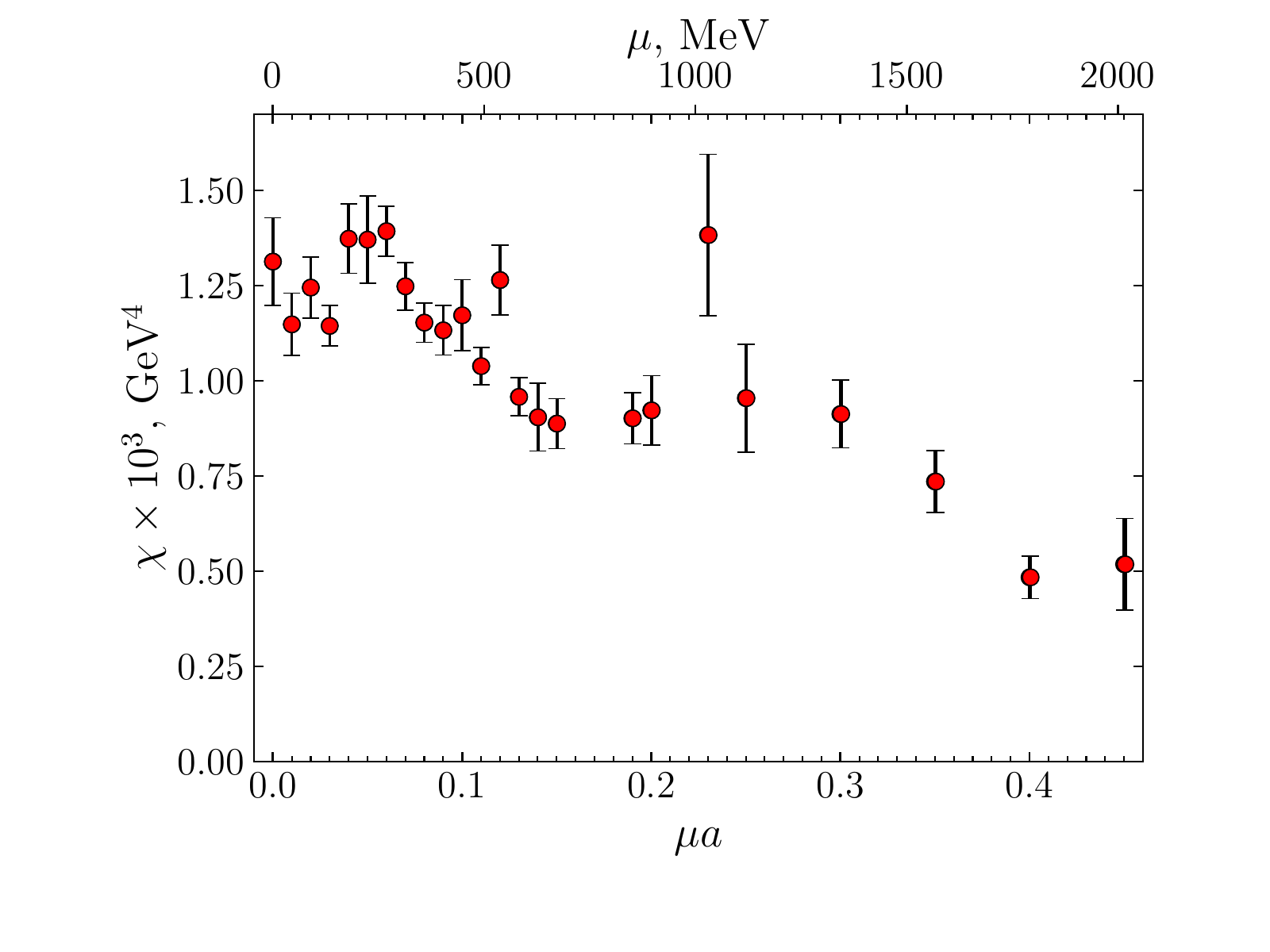}
    \caption{Topological susceptibility in energy units, scaled by $10^3$ for the better visual presentation, as a function of the chemical potential.}
    \label{fig:top_susceptibility}
\end{figure}

To study how nonzero baryon density influences the topological properties of QC$_2$D, we calculated the topological susceptibility $\chi$ for various values of the baryon chemical potentials under study. The result of this calculation is presented in Fig.~\ref{fig:top_susceptibility}. Despite large uncertainties at a few points, it may be seen from this plot that the topological susceptibility slowly decreases with rising of the chemical potential. This result is in disagreement with one recent study~\cite{Iida:2019rah}, but it agrees with the results of the papers~\cite{Alles:2006ea, Alles:2006re}. We believe that our findings on the topological susceptibility are supported by the other results of this paper. In particular, in this section we observed, that chromoelectric fields are screened in dense matter. For this reason one can expect that topological fluctuations are suppressed by dense matter as compared to vacuum.

\section{Equation of state of dense QC$_2$D}
\label{sec:eos}

To study the EoS of dense two-color matter in this paper we are going to use equations (\ref{pressure})-(\ref{eq:entropy}). For the functions $\beta(g)$ the two-loop perturbative expression is used, which is independent on regularization:
\beq
\beta(g) = - \frac {8} {g^2} \biggl ( 3 \frac {g^2} {4\pi^2} + 29 \frac {g^4} {64\pi^4} \biggr )\,.
\label{beta_function}
\eeq
For the function $\gamma(g)$ in the calculation we use the perturbative one-loop expression (\ref{gamma_function}), which similarly to (\ref{beta_function}) does not depend on regularization
\beq
\gamma(g)=1+ \frac 9 {32 \pi^2} g^2\,.
\label{gamma_function}
\eeq
Notice, that the use of the one-loop expression might lead to systematic uncertainty. However, we believe that we are close to the continuum limit and the expression (\ref{gamma_function}) is a good approximation for the actual $\gamma(g)$ function. This statement is supported by the findings in~\cite{Cheng:2007jq} devoted to the EoS of $SU(3)$ QCD. In this paper it was found, that for sufficiently small $g^2$ the $\gamma(g)$ is well described by one-loop formula similar to (\ref{gamma_function}). The fact that we work close to the continuum limit can be seen from the following observation: if instead of the two-loop $\beta(g)$ function one uses one-loop expression the results for the $\beta(g)$ will change by 10\,\%. Moreover, the $\gamma(g)$ function enters to the fermion contribution to the anomaly. From what follows it will be clear, that the fermion contribution to the anomaly is quite small and it modifies the EoS within the uncertainty of the calculation.

\begin{figure}[t]
    \centering
    \includegraphics[scale=0.6]{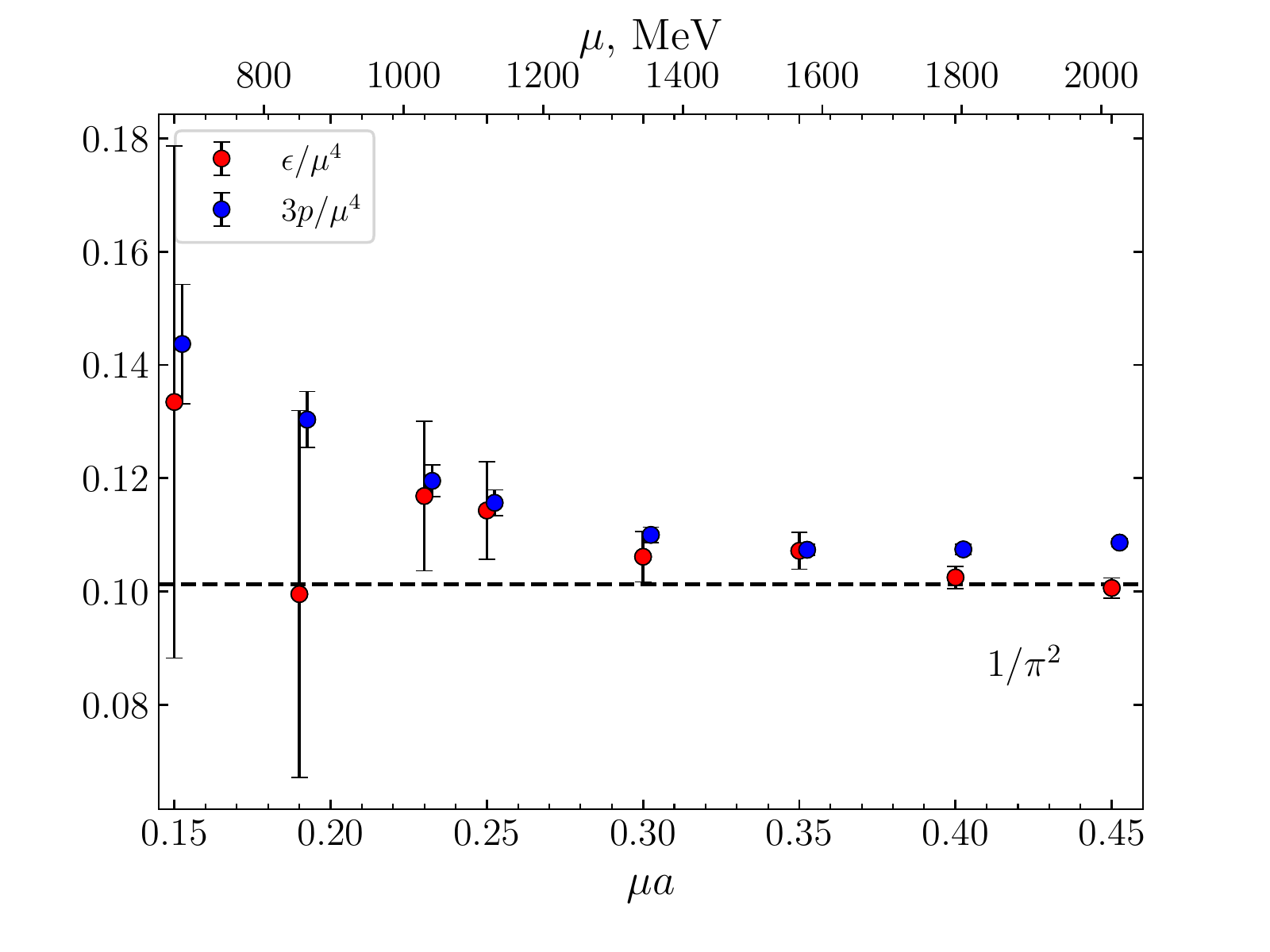}
    \caption{The energy density and the pressure divided by $\mu^4$ as a function of chemical potential (blue circles are slightly shifted for the better visibility). The dashed line corresponds to the $\epsilon$ and $3p$ of a free relativistic quark gas $\epsilon=3p=\mu^4/\pi^2$.}
    \label{fig:energy_pressure}
\end{figure}

\begin{figure}[t]
    \centering
    \includegraphics[scale=0.6]{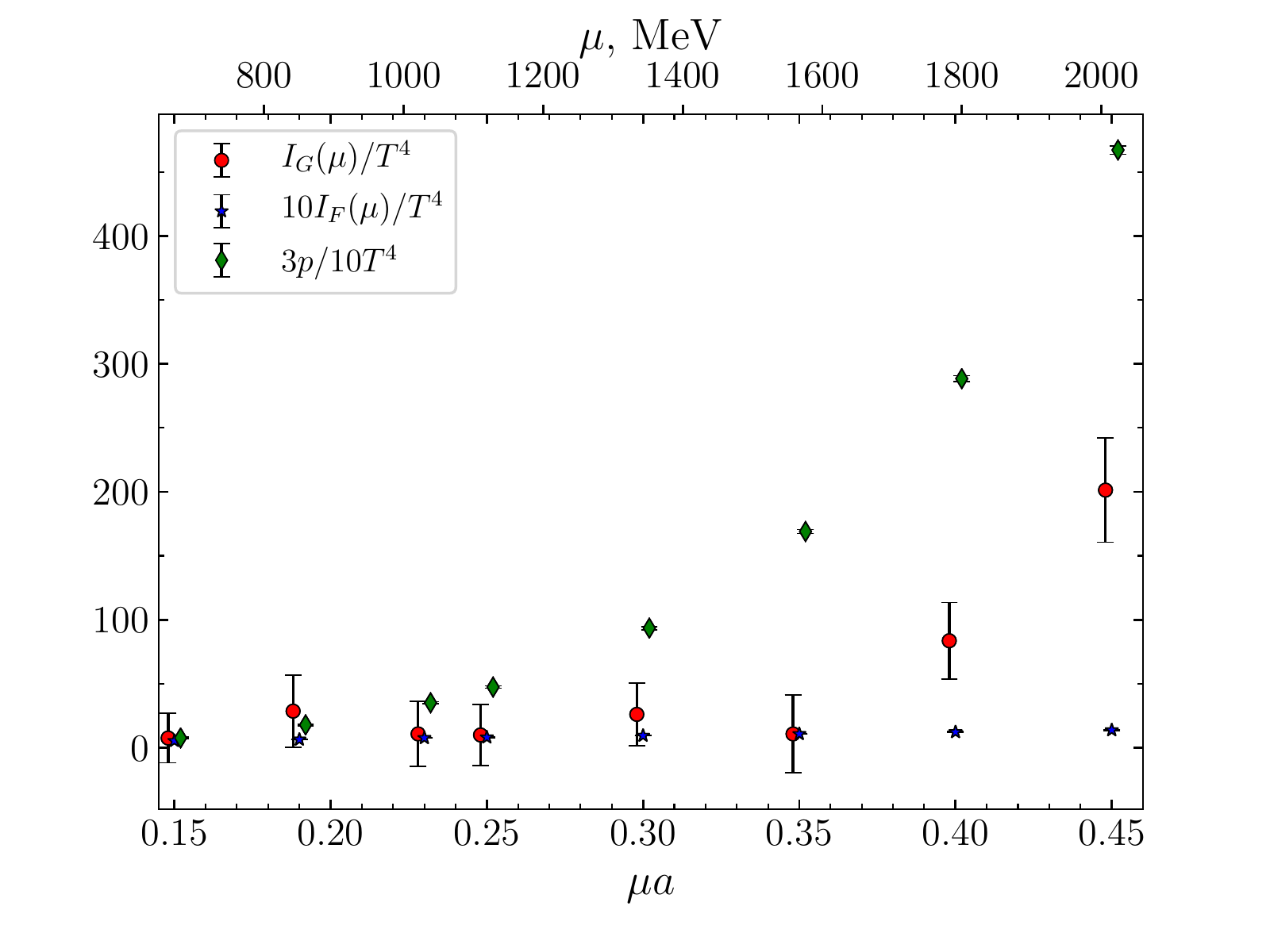}
    \caption{The gluon $I_G$ and fermion $I_F$ contributions to the anomaly, defined in (\ref{eq:I_G}) and (\ref{eq:I_F}) respectively, and pressure $p$ as functions of the chemical potential. In order to plot these observables in one figure we rescaled them.
    }
    \label{fig:anomaly}
\end{figure}

In the Fig.~\ref{fig:anomaly} we plot the gluon $I_G$ and fermion $I_F$ contributions to the anomaly and the pressure $p$ as a function of the chemical potential. In order to plot these observables in one figure we have rescaled them. The energy density (\ref{eq:energy_density}) is the sum of the $I_G$, $I_F$ and $3p$, thus using Fig.~\ref{fig:anomaly} one can explore the role of these three contributions in the EoS. From the Fig.~\ref{fig:anomaly} it may be seen that the smallest contribution is the fermion part of the anomaly, $I_F$. The next (by the size) is the gluon contribution to the anomaly, $I_G$. Unfortunately, the uncertainty of the calculation of this observable is quite large, and the $I_G$ develops nonzero values only in the region $a\mu \geq 0.4$. Notice that for all the values of $\mu$ under study the fermionic contribution $I_F$ is much smaller than the uncertainty of the calculation of the gluonic contribution $I_G$. The term $3p$ provides the largest contribution to the energy density (\ref{eq:energy_density}). In the region, which can be well described by ChPT, $a\mu < 0.12$, the $3p$ term is compatible to the uncertainty in $I_G$. In the following region of the phase diagram, where the system becomes dense, $a\mu > 0.12$, the $3p$ term is larger than the gluonic contribution $I_G$, but the uncertainty in the energy density remains quite large. Finally, in the BCS phase the uncertainty in the energy density $\epsilon$ becomes small.

In Fig.~\ref{fig:energy_pressure} we present the energy density and the pressure, divided by $\mu^4$, as functions of the chemical potential. The dashed line corresponds to the EoS of free relativistic quark gas $\epsilon = 3p = \mu^4/\pi^2$. It may be observed from the Fig.~\ref{fig:energy_pressure}, that in the BCS phase the EoS is well described by the EoS of free relativistic quarks. In our simulations the quark mass is quite large leading to large pion mass, but nonzero quark mass does not play an important role in the BCS phase.

Now let us focus on the entropy density (\ref{eq:entropy}). According to the third law of thermodynamics the entropy approaches to some constant value as temperature approaches to zero. If the ground state is not degenerate, the entropy is zero. In our simulations due to nonzero value of the $\lambda$-parameter there is no degeneracy in the system under study, thus one can expect that the entropy is zero. Our results confirm that $s = 0$ within the uncertainty of the calculation for all values of the chemical potential under consideration. 

Previously the EoS of dense QC$_2$D was studied in the papers~\cite{Cotter:2012mb, Boz:2019enj}, where lattice simulation was carried out with dynamical Wilson fermions. It is rather difficult to compare our results and the results obtained there due to large uncertainties of the calculation at small values of chemical potential. However, in the BCS phase the EoS from~\cite{Cotter:2012mb, Boz:2019enj} is well described by the EoS of free relativistic quarks, which agrees with the results of the present paper.

\begin{figure}[t]
	\includegraphics[scale=0.5]{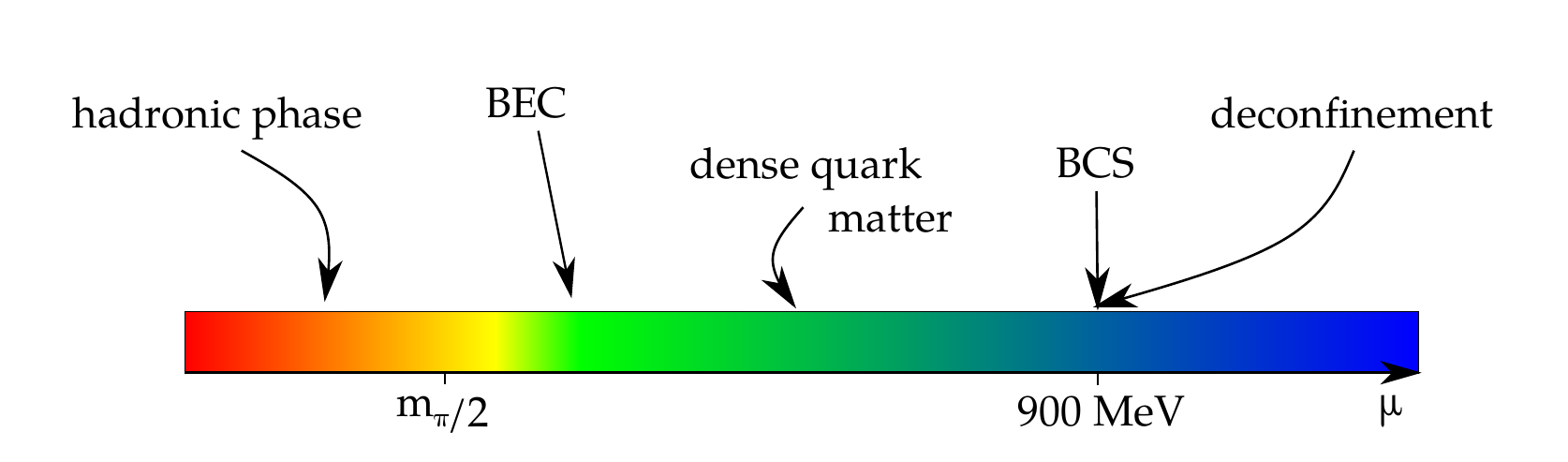}
    \caption{Schematic phase diagram of dense two-color QCD at low temperatures.}
    \label{fig:phase_diagram}
\end{figure}

\section{Discussion and conclusion}
\label{sec:phasediag}

In this paper, we carried out lattice study of the phase diagram of dense two-color QCD with $N_f = 2$ quarks and thermodynamic properties of this system. This study was conducted at low temperature and for the baryon chemical potential in the region $\mu \in (0,\,2000)$\,MeV.

The results of this and previous~\cite{Braguta:2016cpw, Bornyakov:2017txe, Astrakhantsev:2018uzd} studies suggest the following phase structure of dense two-color QCD at low temperatures (a schematic phase diagram of dense two-color QCD is shown in Fig.~\ref{fig:phase_diagram}): for small values of chemical potential ($\mu<m_{\pi}/2$) the system is in the hadronic phase, and the chiral symmetry is broken; at $\mu=m_{\pi}/2=371(8)$\,MeV there is a transition to a
phase, where scalar diquarks form a Bose-Einstein condensate, diquark condensate and baryon density develop nonzero values. In the massless limit there is no chiral symmetry breaking, if the diquarks are condensed. However, for massive quarks the chiral condensate is finite, proportional to the quark mass and decreases with increasing chemical potential.

In the ChPT the interactions between different degrees of freedom are accounted for by perturbation theory, so they are assumed to be weak. Together with the fact that in two-color QCD the diquarks are baryons one may state, that the system on the right side of the phase transition at $\mu \geq m_{\pi}/2$, but not at too large chemical potential, is similar to a dilute baryon gas. Lattice results for the baryon density, diquark and chiral condensates are well described by ChPT up to $\mu < 540$\,MeV.

Increasing the baryon density further, we proceed to dense matter, where the interactions between baryons cannot be fully accounted within perturbation theory. This transition manifests itself in terms of the deviation of different observables from the ChPT predictions. In particular, in this paper the deviation is well pronounced in the diquark condensate and the baryon density.

At sufficiently large baryon density ($\mu \sim 900$\,MeV, $a\mu \sim 0.20$) some observables of the system under study can be described using Bardeen-Cooper-Schrieffer theory (BCS phase). In particular, the baryon density is well described by the density of non-interacting fermions which occupy a Fermi sphere of radius $r_F=\mu$. Moreover, the diquark 
condensate, which plays the role of a condensate of Cooper pairs, is proportional to the Fermi surface.

In the region $a\mu < 0.2$ the system under study is in the confinement phase. However, at $\mu \sim 900$\,MeV ($a\mu \sim 0.2$) we observe confinement/deconfinement transition in dense two-color QCD~\cite{Bornyakov:2017txe}. This transition manifests itself in a rise of the Polyakov loop and vanishing of the string tension. It was also found that, after deconfinement is achieved, spatial string tension $\sigma_s$ decreases monotonically and ends up vanishing at $\mu_q \geq 2000$\,MeV ($a\mu \geq 0.45$). Notice, however, that this region is spoiled by lattice artifacts.

In addition to the phase diagram we have studied how nonzero baryon density effects the gluon background. We found that chromoelectric field decreases with rising of baryon density. We believe that this behaviour can be attributed to well known Debye screening of chromoelectric field in dense matter. This phenomenon was also observed in the study of Polyakov loops correlators~\cite{Astrakhantsev:2018uzd} and gluon propagators~\cite{Bornyakov:2019jfz, Bornyakov:2020kyz} in dense matter.
As for the chromomagnetic field, within the uncertainty it remains the same as compared to its vacuum value up to the chemical potential $a\mu \sim 0.2$, then in the region $a\mu > 0.2$ magnetic field increases with rising of the baryon density. This behaviour can be explained if we recall, that magnetic screening in QCD matter is related to nonperturbative spatial confinement. In the paper~\cite{Bornyakov:2017txe} it was found, that in the region $a\mu> 0.2$ the spatial string tension decreases, {\it i.e.} spatial confinement plays less important role. As the result chromomagnetic field is less screened. Similar results were obtained in papers~\cite{Bornyakov:2019jfz, Bornyakov:2020kyz}.

To study how nonzero baryon density influences the topological properties of QC$_2$D we have calculated the topological susceptibility $\chi$. The topological susceptibility slowly decreases with rising of the chemical potential. We believe, that this decrease of $\chi$ with increasing chemical potential is related to the screening of chromoelectric fields in dense matter. 

Finally in this paper the equation of state of dense QC$_2$D was studied. In particular, we calculated the pressure, the trace anomaly and the energy density. Although it is possible to calculate the pressure with rather good accuracy, the uncertainty in the trace anomaly is rather large. As a result, good accuracy in the energy density can be achieved only in the BCS phase. It is interesting to note, that in the BCS phase the equation of state is well described by the corresponding formulae for free relativistic fermions $\epsilon \simeq 3p,\,p \simeq \mu^4/3\pi^2$. The entropy density remains zero within the uncertainty of the calculation for all values of the chemical potential as it should be in the $\lambda \neq 0$ case. 

At the end of this paper we are going to discuss the dependence of results on the diquark source $\lambda$. Within the accuracy of the calculation we don't see any $\lambda$-dependence of the gluonic observables. As for the fermionic observables in the region of the phase diagram, which is well described by the ChPT, there is $\lambda$-dependence. In particular, both the quark number density and the diquark condensate are zero for $\lambda = 0$ in the region $\mu<m_{\pi}/2$, and finite value of the $\lambda$-parameter leads to nonzero values of these observables. Close to the phase transition $\mu \sim m_{\pi}/2$ there is also $\lambda$-dependence of the fermionic observables. However, when one moves from the region of the phase transition $\mu \sim m_{\pi}/2$ further to dense matter, the $\lambda$-dependence becomes weak and the fermionic observables at $\lambda/m \sim 0.1$ are close to their values in the limit $\lambda \to 0$. This result is in agreement with the previous findings in~ \cite{Braguta:2016cpw, Wilhelm:2019fvp}. The $\lambda$-dependence of the EoS is related to the mentioned $\lambda$-dependence of the quark number density, because the chiral condensate does not depend on it. In the region $\mu<m_{\pi}/2$ and in the vicinity of the phase transition $\mu \sim m_{\pi}/2$ the quark number density is small and does not contribute significantly to the EoS, and in the BCS phase its dependence on $\lambda$ is weak, thus the $\lambda$-dependence of the EoS is weak.

\section*{Acknowledgments}

We would like to thank Prof. Naoki Yamamoto for useful discussions. The work of V.\,V.\,B., which consisted of study of the equation of state and physical interpretation of lattice data, was supported by grant from the Russian Science Foundation (project number 16-12-10059). N.\,Yu.\,A.,  A.\,Yu.\,K. and A.\,A.\,N. acknowledge the support of the Russian Foundation for Basic Research (project no. 18-32-20172 mol\_a\_ved). A.\,A.\,N. is also grateful for the support from STFC grant ST/P00055X/1. This work has been carried out using computing resources of the federal collective usage center Complex for Simulation and Data Processing for Mega-science Facilities at NRC ``Kurchatov Institute'',~\url{http://ckp.nrcki.ru/}. In addition, the authors used the cluster of the Institute for Theoretical and Experimental Physics and the supercomputer of Joint Institute for Nuclear Research ``Govorun''.

\bibliography{refs}

\end{document}